\documentclass[epj]{webofc-mod}
\usepackage[varg]{txfonts}   % Web of Conferences font
\usepackage[hyperindex,breaklinks]{hyperref}
\usepackage{url} 

\def\to{\ensuremath{\rightarrow}}

\DeclareGraphicsExtensions{.pdf,.png}
\wocname{EPJ Web of Conferences}
\woctitle{ICNFP 2016}
\begin{document}
\selectlanguage{english}
\hfill KCL-PH-TH/2016-70

\hfill IFIC/16-93
\title{Physics reach of MoEDAL at LHC: magnetic monopoles, \\ supersymmetry and beyond}

\author{Nick E.\ Mavromatos\inst{1}\fnsep\thanks{\email{nikolaos.mavromatos@kcl.ac.uk}} \and
        Vasiliki A.\ Mitsou\inst{2}\fnsep\thanks{\email{vasiliki.mitsou@ific.uv.es}} 
\\ On behalf of the MoEDAL Collaboration} 

\institute{Theoretical Particle Physics and Cosmology Group, Department of Physics,
King's College London, Strand, London WC2R 2LS, UK
\and
Instituto de F\'isica Corpuscular (IFIC), CSIC -- Universitat de Val\`encia, 
C/ Catedr\'atico Jos\'e Beltr\'an 2,\\ E-46980 Paterna (Valencia), Spain
}

\abstract{%
MoEDAL is a pioneering experiment designed to search for highly ionising messengers of new physics such as magnetic monopoles or massive (pseudo-)stable charged particles, that are predicted to exist in a plethora of models beyond the Standard Model. Its ground-breaking physics program defines a number of scenarios that yield potentially revolutionary insights into such foundational questions as, are there extra dimensions or new symmetries, what is the mechanism for the generation of mass, does magnetic charge exist, what is the nature of dark matter, and, how did the big-bang develop at the earliest times. MoEDAL's purpose is to meet such far-reaching challenges at the frontier of the field. The physics reach of the existing MoEDAL detector is discussed, giving emphasis on searches for magnetic monopoles, supersymmetric (semi)stable partners, doubly charged Higgs bosons, and exotic structures such as black-hole remnants in models with large extra spatial dimensions and D-matter in some brane theories.
}
\maketitle
%
%%%%%%%%%%%%%%%%%%%%%%%%%%%%%%%%%%%%%%%%%%%%%%%%%%%%%%%%%%%%%%%%%%%%%%%%%%%%%%%%%%%%%%%%%%%%%%%%%%%%
%%%%%%%%%%%%%%%%%%%%%%%%%%%%%%%%%%%%%%%%%%%%%%%%%%%%%%%%%%%%%%%%%%%%%%%%%%%%%%%%%%%%%%%%%%%%%%%%%%%%
\section{Introduction}\label{sc:intro}

MoEDAL (Monopole and Exotics Detector at the LHC)~\cite{moedal-web,moedal-tdr,jim}, the $7^{\rm th}$ experiment at the Large Hadron Collider (LHC)~\cite{LHC}, was approved by the CERN Research Board in 2010. It is designed to search for manifestations of new physics through highly-ionising particles in a manner complementary to ATLAS and CMS~\cite{DeRoeck:2011aa}. The most important motivation for the MoEDAL experiment is to pursue the quest for magnetic monopoles and dyons at LHC energies. Nonetheless the experiment is also designed to search for any massive, stable or long-lived, slow-moving particles~\cite{Fairbairn07} with single or multiple electric charges arising in many scenarios of physics beyond the Standard Model (SM). A selection of the physics goals and their relevance to the MoEDAL experiment are described here and elsewhere~\cite{creta2014}. For an extended and detailed account of the MoEDAL discovery potential, the reader is referred to the recently published \emph{MoEDAL Physics Review}~\cite{Acharya:2014nyr}.

The structure of this paper is as follows. Section~\ref{sc:detector} provides a brief description of the MoEDAL detector. The physics reach of MoEDAL as far as magnetic monopoles and monopolia is discussed in Section~\ref{sc:mm}, whilst Section~\ref{sc:susy} is dedicated to supersymmetric models predicting massive (meta)stable states. Scenarios with doubly-charged Higgs bosons and their observability in MoEDAL are highlighted in Section~\ref{sc:lrsm}. Highly-ionising exotic structures in models with extra spatial dimensions, namely microscopic black holes and D-matter, relevant to MoEDAL are briefly mentioned in Sections~\ref{sc:bh} and Sections~\ref{sc:dmatter}, respectively. The paper concludes with a summary and an outlook in Section~\ref{sc:summary}.

%%%%%%%%%%%%%%%%%%%%%%%%%%%%%%%%%%%%%%%%%%%%%%%%%%%%%%%%%%%%%%%%%%%%%%%%%%%%%%%%%%%%%%%%%%%%%%%%%%%%
%%%%%%%%%%%%%%%%%%%%%%%%%%%%%%%%%%%%%%%%%%%%%%%%%%%%%%%%%%%%%%%%%%%%%%%%%%%%%%%%%%%%%%%%%%%%%%%%%%%%
\section{The MoEDAL detector}\label{sc:detector}

\begin{figure}[htb]
\begin{center}
\includegraphics[width=0.58\textwidth]{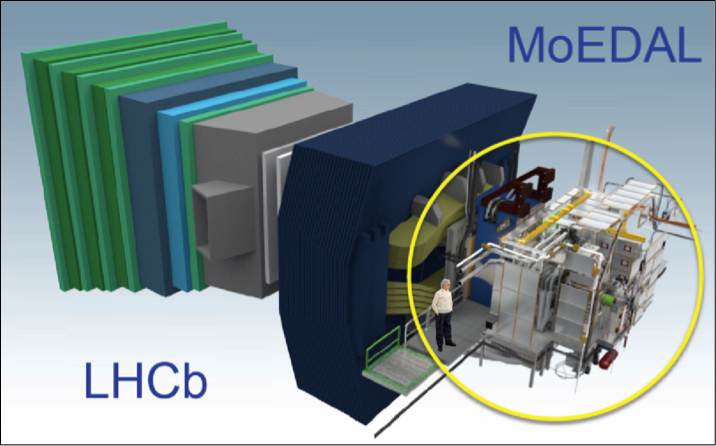}
\caption{ A three-dimensional schematic view of the MoEDAL detector (on the right) around the LHCb VELO region at Point~8 of the LHC.}
\label{Fig:moedal-lhcb}
\end{center}
\end{figure}

The MoEDAL detector~\cite{moedal-tdr} is deployed around the intersection region at Point~8 of the LHC in the LHCb experiment Vertex Locator (VELO)~\cite{LHCb-detector} cavern. A three-dimensional depiction of the MoEDAL experiment is presented in Fig.~\ref{Fig:moedal-lhcb}. It is a unique and largely passive LHC detector comprised of four sub-detector systems. 

The main sub-detector system is made of a large array of CR39\textregistered,  Makrofol\textregistered\ and Lexan\textregistered\ nuclear track detector (NTD) stacks surrounding the intersection area. The passage of a highly-ionising particle through the plastic detector is marked by an invisible damage zone along the trajectory. The damage zone is revealed as a cone-shaped etch-pit when the plastic detector is etched using a hot sodium hydroxide solution. Then the sheets of plastics are scanned looking for aligned etch pits in multiple sheets. The MoEDAL NTDs have a threshold of $Z/\beta\sim5$, where $Z$ is the charge and $\beta=v/c$ the velocity of the incident particle. In proton-proton collision running, the only source of known particles that are highly ionising enough to leave a track in MoEDAL NTDs are spallation products with range that is typically much less than the thickness of one sheet of the NTD stack. In that case the ionising signature will be that of a very low-energy electrically-charged \emph{stopped} particle. This signature is distinct to that of a  \emph{penetrating} electrically or magnetically charged particle that will usually traverse every sheet in a MoEDAL NTD stack, accurately demarcating a track that points back to the collision point. During the heavy-ion running one might expect a background from high ionising fragments, which however are produced in the far forward direction and do not fall into the acceptance of the MoEDAL detector.

A unique feature of the MoEDAL detector is the use of paramagnetic magnetic monopole trappers (MMTs) to capture electrically- and magnetically-charged highly-ionising particles. The aluminium absorbers of MMTs will be subject to an analysis looking for magnetically-charged particles at a remote magnetometer facility~\cite{Joergensen:2012gy,DeRoeck:2012wua}. The search for the decays of long-lived electrically charged particles that are stopped in the trapping detectors will subsequently be carried out at a remote underground facility.

The only non-passive MoEDAL sub-detector system comprises an array of several TimePix pixel devices that form a real-time radiation monitoring system dedicated to the monitoring of highly-ionising background sources in the MoEDAL cavern.

%%%%%%%%%%%%%%%%%%%%%%%%%%%%%%%%%%%%%%%%%%%%%%%%%%%%%%%%%%%%%%%%%%%%%%%%%%%%%%%%%%%%%%%%%%%%%%%%%%%%
%%%%%%%%%%%%%%%%%%%%%%%%%%%%%%%%%%%%%%%%%%%%%%%%%%%%%%%%%%%%%%%%%%%%%%%%%%%%%%%%%%%%%%%%%%%%%%%%%%%%
\section{Magnetic monopoles}\label{sc:mm}

The MoEDAL detector is designed to fully exploit the energy-loss mechanisms of magnetically charged particles~\cite{Dirac1931kp,Diracs_idea,tHooft-Polyakov,mm,Cho1996qd}  in order to optimise its potential to discover these messengers of new physics. There are various theoretical scenarios in which magnetic charge would be produced  at the LHC~\cite{Acharya:2014nyr}: (light) 't Hooft-Polyakov monopoles~\cite{tHooft-Polyakov,Vento2013jua}, electroweak monopoles~\cite{Cho1996qd} and monopolium~\cite{Diracs_idea,khlopov,Monopolium,Monopolium1}. Magnetic monopoles that carry a non-zero magnetic charge and dyons possessing both magnetic and electric charge are among the most fascinating  hypothetical particles. Even though there is no generally acknowledged empirical  evidence for their existence, there are strong theoretical reasons to believe that they do exist, and they are predicted by many theories including grand unified theories and superstring theory~\cite{Rajantie:2012xh,rajantiept}. 
 
%%%%

The theoretical motivation behind the introduction of magnetic monopoles is the symmetrisation of the Maxwell's equations and the explanation of the charge quantisation~\cite{Dirac1931kp}. Dirac showed that the mere existence of a monopole in the universe could offer an explanation of the discrete nature of the electric charge, leading to the Dirac Quantisation Condition (DQC),

\begin{equation} \alpha~  g = \frac{N}{2} e , \quad  N = 1, 2, ... , 
\label{eq:dqc}\end{equation}

\noindent where $e$ is the electron charge, $\alpha = \frac{e^2}{4\pi \hbar\, c \varepsilon_0 } = \frac{1}{137}$ is the fine structure constant (at zero energy, as appropriate to the fact that the quantization condition of Dirac pertains to long (infrared) distances from the centre of the monopole), $\varepsilon_0$ is the vacuum permitivity, and $g$ is the monopole magnetic charge. In Dirac's formulation, magnetic monopoles are assumed to exist as point-like particles and quantum mechanical consistency conditions lead to Eq.~(\ref{eq:dqc}), establishing the value of their magnetic charge. Although monopoles symmetrise Maxwell's equations in form, there is a numerical asymmetry arising from the DQC, namely that the basic magnetic charge is much larger than the smallest electric charge. A magnetic monopole with a single Dirac charge ($g_{\rm D}$) has an equivalent electric charge of  $\beta(137e/2)$. Thus  for a relativistic monopole the energy loss is around $4,\!700$ times ($68.5^2$) that of a minimum-ionising electrically-charged particle. The monopole mass remains a free parameter of the theory.

%%%%
%%%%%%%%%%%%%%%%%%%%%%%%%%%%%%%%%%%%%%%%%%%%%%%%%%%%%%%%%%%%%%%%%%%%%%%%%%%%%%%%%%%%%%%%%%%%%%%%%%%%
\subsection{Electroweak monopoles}\label{sc:cho}

The original monopole solution of `t Hooft and Polyakov~\cite{tHooft-Polyakov} pertain to a simple compact (from the point of view of its topology in internal space) gauge group, such as $SU(2)$ or the more phenomenologically realistic Grand Unified Theory (GUT) group $SU(5)$. An important ingredient of the monopole in such models is played by the scalar (spontaneous symmetry breaking Higgs-like) sector of the model. The monopole is an extended object, with complicated structure, with a spherical core of radius $R_c$. inside which the scalar fields acquires a zero vacuum expectation value, $\langle \vec \phi \rangle = 0$, while outside the fields approach their non-trivial v.e.v. $\langle \vec \phi \rangle = \eta \ne 0 $ asymptotically (where $\vec \phi$ denotes the appropriate field multiplet).

The mass of such monopoles are finite. In the Georgi-Glashow model of spontaneous symmetry breaking analysed by `t Hooft~\cite{tHooft-Polyakov}, based on the $SU(2)$ gauge group (with a Higgs triplet), the mass of the monopole may be of order of the electroweak scale. In GUT gauge group cases, the monopole mass is of 
order of the GUT scale, and thus outside of the LHC reach. The quantisation condition for the magnetic charge of the monopole in such theories arises because of the non trivial homotopy groups of the pertinent gauge group. In the case, e.g., of the $SU(2)$ group in the Georgi-Glashow model, the internal-space sphere $S^2$ winds around a configuration space sphere a number of times, defining different topological sectors, characterised by discrete monopole charges.

From this point of view, the Standard Model, with a Higgs doublet of fields and a gauge group $SU(2) \otimes U_Y(1)$, cannot support monopole solutions because its 
gauge group is not a simple one. However, Cho and Maison~\cite{Cho1996qd} argued that the presence of $U_Y(1)$ is important in providing ${\rm CP}^1$ structures by considering the quotient group $SU(2) \otimes U_Y(1) / U_{\rm em}(1) $, where $U_{\rm em}(1)$ is the (unbroken) group of electromagnetism, thus admitting monopole solutions as a result of the fact that $\pi_2({\rm CP}^1) = n \in Z$, with the standard Higgs doublet as a CP$^1$ field. 
This monopole solution unfortunately had an energy functional which was singular at the origin $r \to 0$. Because of this singularity, the monopole mass could not be determined without embedding the theory into a larger group. It could also be possible that although such solutions are topologically allowed in the Standard Model, nevertheless because of this infinite energy are strongly coupled and bound into finite energy monopole-antimonopole pairs (``monopolia'') which are not highly ionising. Such states, although not relevant directly for MoeDAL physics, may be constrained by means of other indirect methods at colliders, which we shall not discuss here. 

Cho and collaborators have recently suggested~\cite{cho2} that a finite energy Cho-Maison monopole, but with magnetic charge twice the elementary Dirac charge $2g_D \equiv 137\, e$, 
 can be a solution of a modified Standard Model Langrangian with a non-trivial vacuum permitivity (gauge invariant) function of the Higgs field modulus $\phi^\dagger \phi$ in front of the kinetic term of the hypercharge $U_Y(1)$ gauge field:
\begin{equation}\label{u1term}
\mathcal L \ni -\frac{1}{4} \varepsilon (\phi^\dagger \phi) \, \mathcal G_{\mu\nu}\, \mathcal G^{\mu\nu}~,
\end{equation}
where $\mathcal G_{\mu\nu}$ denotes the $U_Y(1)$ gauge potential field strength, and the function $\varepsilon  (\phi^\dagger \phi)$ is such that far away from the monopole centre approaches one, i.e.\ $\varepsilon  (\phi^\dagger \phi) \to 1, r \to \infty$, while as $r \to 0$ vanishes in such a way that the energy of the monopole is finite. 
The term (\ref{u1term}) ensures that there is an effective running $U_Y(1)$ coupling $\overline g^\prime = g^\prime /\varepsilon $, with $g^\prime$ the Standard Model hypercharge coupling.
Simple functions $\epsilon(\phi^\dagger \phi)$ with such properties are~\cite{cho2} 
\begin{equation}
\varepsilon(\phi^\dagger\, \phi) = \Big|\frac{\phi^\dagger \, \phi}{\eta^2}\Big |^{N/2}, \quad N \ge 4 + 2 \sqrt{3} \simeq 7.46 ~,
\end{equation}
with $\eta$ the (real) v.e.v.\ of $\phi$ in the symmetry broken phase. Unfortunately the LHC phenomenology of such simple models excludes them, in the sense that they can produce 
unacceptably high signals of $H \to \gamma\gamma$, where $H$ is the Higgs field, defined as $\phi = \eta + H$ in the symmetry broken phase, i.e.\ at LHC energies and hence for distances far away from the monopole core.  Expanding in such a case $\varepsilon $ in powers of $H \ll \eta$,  one obtains from (\ref{u1term}) dimension-six operators in an effective field theory framework~\cite{You}. Applying then the phenomenological analysis of Ref.~\cite{sanz}, one observes that 
acceptable LHC phenomenology of the modified standard model can be obtained for more complicated $\varepsilon(\phi^\dagger \phi)$ functions~\cite{You}, which consist of algebraic combinations of various powers of the Higgs modulus yielding acceptably suppressed rates of $H \to \gamma\gamma$. 
Such constructions can yield monopole masses at least of order 5 TeV~\cite{You}, which makes the case interesting for MoeDAL~\footnote{Self-gravitating monopole solutions have been considered in this context in Ref.~\cite{chograv}, with the conclusion, as expected, that gravity has the tendency to reduce the monopole mass. However the order of the mass remains the same as in the flat spacetime case above.}. Unfortunately at present there is no microscopic justification of such modifications, nevertheless from a phenomenological viewpoint 
such models can be falsified at MoEDAL or other LHC experiments, given that they can in principle allow for the production of highly ionising monopole/antimonopole pairs from proton-proton collisions. The extended nature of the respective monopole solutions, however, most probably imply strong suppression factors in the corresponding cross sections, which make such processes rare.

%%%%%%%%%%%%%%%%%%%%%%%%%%%%%%%%%%%%%%%%%%%%%%%%%%%%%%%%%%%%%%%%%%%%%%%%%%%%%%%%%%%%%%%%%%%%%%%%%%%%
\subsection{Global $O$(3) monopoles}\label{sc:global}

Global monopoles have been proposed~\cite{vilenkin} as space time (cosmological) defects allowing for the spontaneous breaking of internal global $SO(3)$ symmetries in non gauged Georgi Glashow models. The appropriate scalar sector of such models consists of triplets of spin zero fields $\chi^a, \, a=1,2,3$, with an appropriate symmetry breaking potential that allows (some of) the fields to acquire v.e.v.\ $\eta$.  There are global (non gauge) monopole solutions in such a model (and its variants), whose stability is still under debate~\cite{debate}, but which may have interesting physics results of relevance to collider physics, given that the associated monopole mass is proportional to $\eta$, ${\mathcal M} \sim \eta$ and the latter can be at TeV scale (if the scalar fields represent somehow new physics fields at TeV scales). The monopole has a core of radius $R_{\rm core} \sim 1/\eta $ and the probability of its production 
in pair with its antimonopole, can be estimated as~\cite{vilenkin,nussinov}
\begin{equation}\label{prob}
\mathcal{P} \propto e^{-4 R_{\rm core}/\lambda_{\rm Compt}} \propto e^{-{\rm const}^\prime /\lambda}, 
\end{equation}
where  $\lambda_{\rm Compt} = 1/\mathcal{M}$ is the Compton wavelength, and $\lambda$ the self-interaction coupling of the scalar $SO(3)$ symmetry breaking sector of the model. 

For large monopoles $R_{\rm core} > \lambda_{\rm Compt}$, and therefore weak self-interaction couplings $\lambda < 1$, a semi-classical situation is reached where the form factor has an exponential suppression and the event is rare. However, there may be models of phenomenological interest in which the coupling $\lambda\gg 1$, is strong. In such cases  
$\mathcal{P}$ is expected to be large, and thus of relevance to collider (including LHC) phenomenology (although  strong coupling can complicate analytical calculations). 
It must be noted, however, that once a monopole/antimonopole pair is produced, energy losses to the massless Goldstone fields associated with the breaking of the global $O(3)$ symmetry, at a rate of order $\eta^2$~\cite{vilenkin} are probably expected, which must be taken into account when considering the relevant phenomenology, as they may lead to the monopole stopping inside the LHC detectors, for instance. 

The global monopoles of Ref.~\cite{vilenkin} are not highly ionising particles, as they carry no magnetic charge. However, as was argued in \cite{vilenkin}, their gravitational effects far away from the monopole centre (i.e. at radial distances $r \to \infty$) are significant, in the sense that the spacetime is not the Minkowski one but has a deficit angle $\eta^2/M_{\rm P}^2$, with $M_{\rm P}$ is the Planck mass. Although minute, such an effect affects~\cite{papav} the forward scattering amplitude of standard model particles that propagate in such backgrounds, leading to 
ring-like angular regions, where the scattering amplitude is very large. The size of such ring-like regions 
is determined by the ratio of the global monopole mass to the Planck mass.
Such peculiar scattering patterns of ordinary standard model particles may indicate indirectly the presence of a neutral global monopole in the detector, despite that fact that the defect is not highly ionising. In this sense, such structures are of interest also to MoEDAL, whose detectors can be sensitive to such phenomena.

Recently~\cite{sarkar} a variant of the global monopole model, including axion fields (with a given ``axion charge'') and a real electromagnetic field, which couples only gravitationally to the scalar $SO(3)$ symmetry breaking sector, has been considered, with the result that the axions 
are capable of inducing electromagnetic monopole solutions with a real magnetic charge of order of the ``axion charge''. Such solutions can of course be highly ionising and thus of direct relevance to MoeDAL, provided their mass is of TeV scale. In such a case both the high ionisation and the peculiar effects~\cite{papav} of the monopole background on the scattering of ordinary particles on them, mentioned in the previous paragraph, are in operation. 

 There are several other models that we shall not list here and which contain some form of monopole or dyon solutions, with masses that could be of order of a few TeV, thus producible in principle at LHC. All such models are extensions of the Standard Model and thus contain new fields in their spectra, that may lead to several signatures of new physics at colliders, beyond the high ionisation induced by the monopoles.  Lacking a concrete theoretical framework, therefore, for magnetic monopoles, an experimental, model independent (as much as possible) search for them, is a pressing need, which can provide the theoretical quest with important guidance, especially if a discovery is made! Fortunately the structure of LHC allow for such searches, and this is the topic of our talk, which will be outlined in the following sections. 
 
%%%%%%%%%%%%%%%%%%%%%%%%%%%%%%%%%%%%%%%%%%%%%%%%%%%%%%%%%%%%%%%%%%%%%%%%%%%%%%%%%%%%%%%%%%%%%%%%%%%%
\subsection{Searches for light monopoles in MoEDAL}\label{sc:lightsearch}

The high ionisation of slow-moving magnetic monopoles and dyons, implies quite characteristic trajectories when such particles interact with the MoEDAL NTDs, which can be revealed during the etching process~\cite{moedal-tdr,Acharya:2014nyr}. In addition, the high magnetic charge of a monopole (which is expected to be at least one Dirac charge $g_D = 68.5 e$ (\emph{cf.} Eq.~(\ref{eq:dqc})) implies a strong magnetic dipole moment, which in turn may result in a strong binding of the monopole with the $^{27}_{13}{\rm Al}$ nuclei of the aluminium MoEDAL MMTs. In such a case, the presence of a monopole trapped in an aluminium bar of an MMT would de detected through the existence of a persistent current, defined as the difference between the currents in the SQUID of a magnetometer before and after the passage of the bar through the sensing coil. 

\begin{figure}[ht]
\begin{center}
  \includegraphics[width=0.48\linewidth]{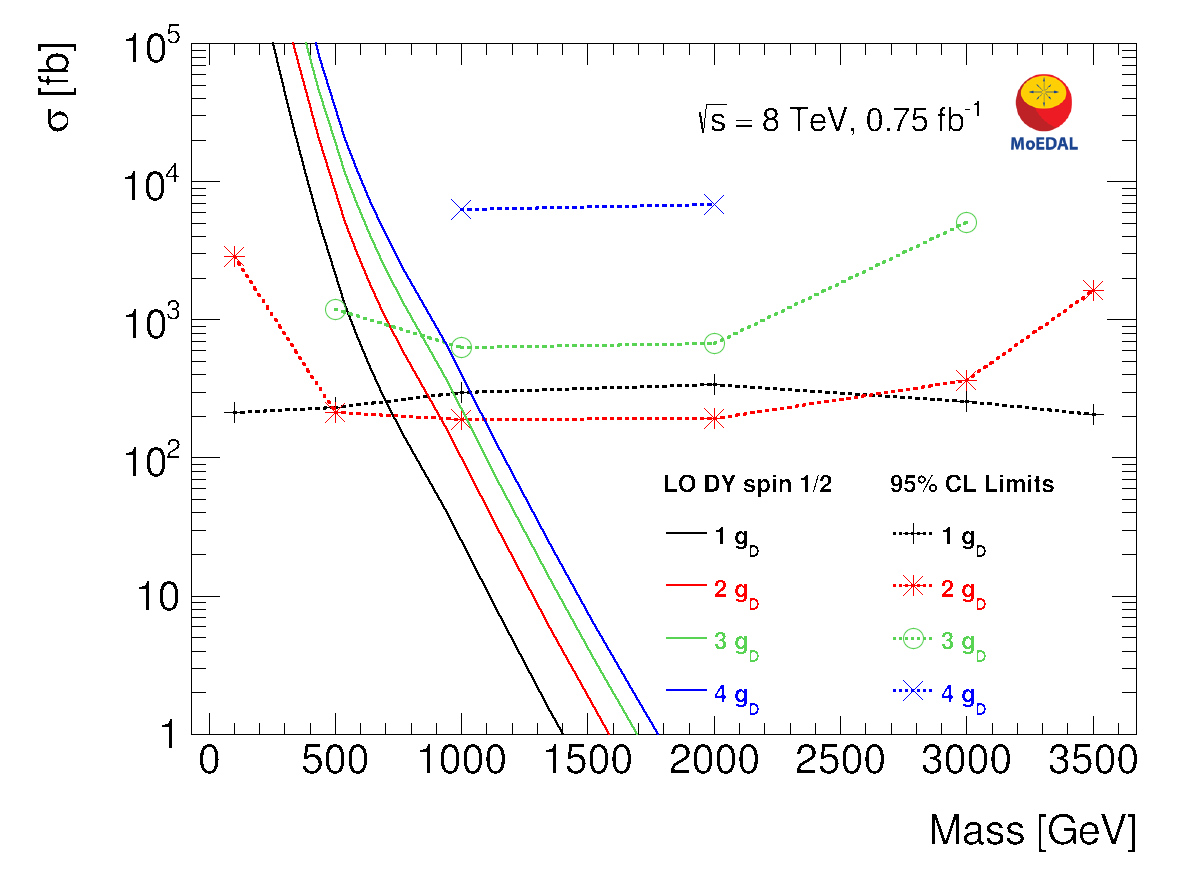}
  \includegraphics[width=0.48\linewidth]{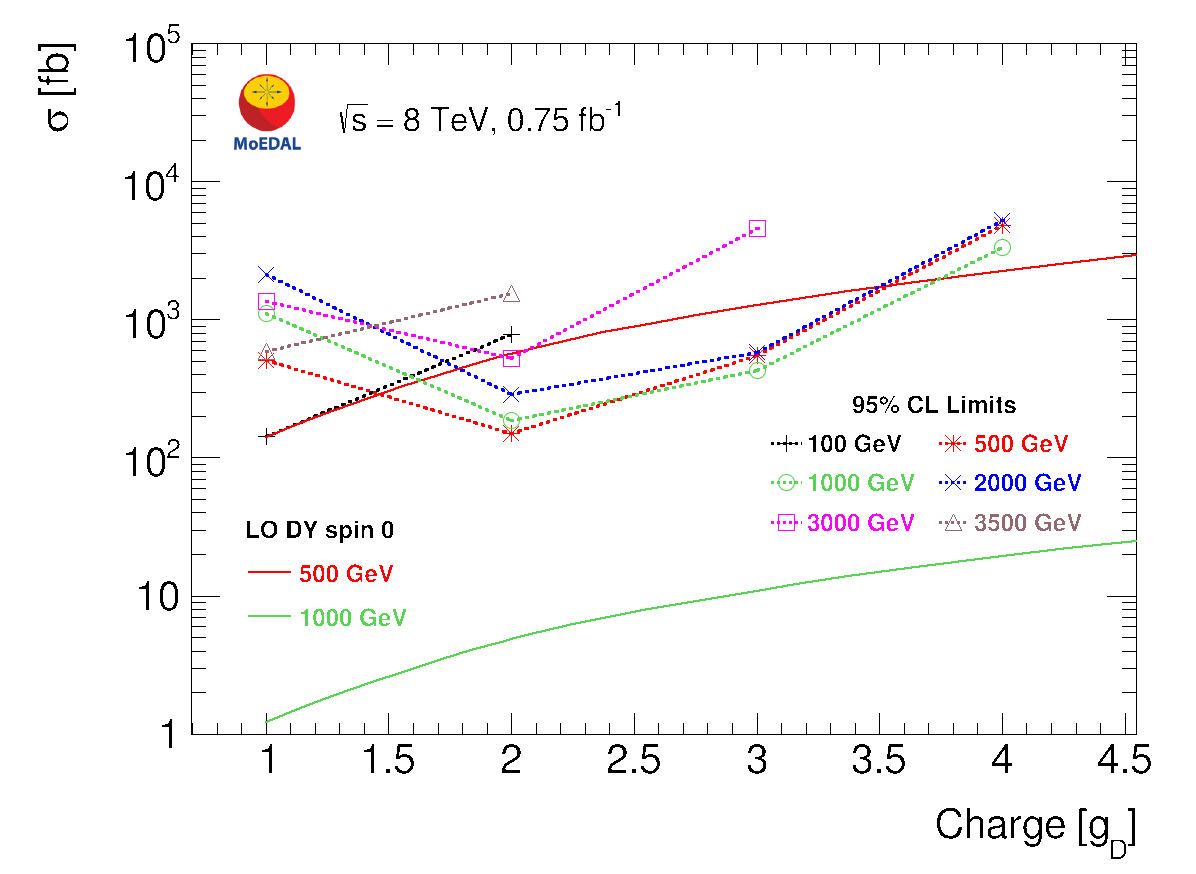}
  \caption{Cross-section upper limits at 95\% confidence level for DY monopole production as a function of mass for spin-1/2 (left) and as a funchion of magnetic charge for spin-0 monopoles (right). The various line styles correspond to different monopole charges (left) or masses (right). The solid lines are DY cross-section calculations at leading order. From Ref~\cite{MMT8TeV}.}
\label{fig:cross_section_limits}
\end{center}
\end{figure}

\begin{figure}[ht]
\begin{center}
  \includegraphics[width=0.58\linewidth]{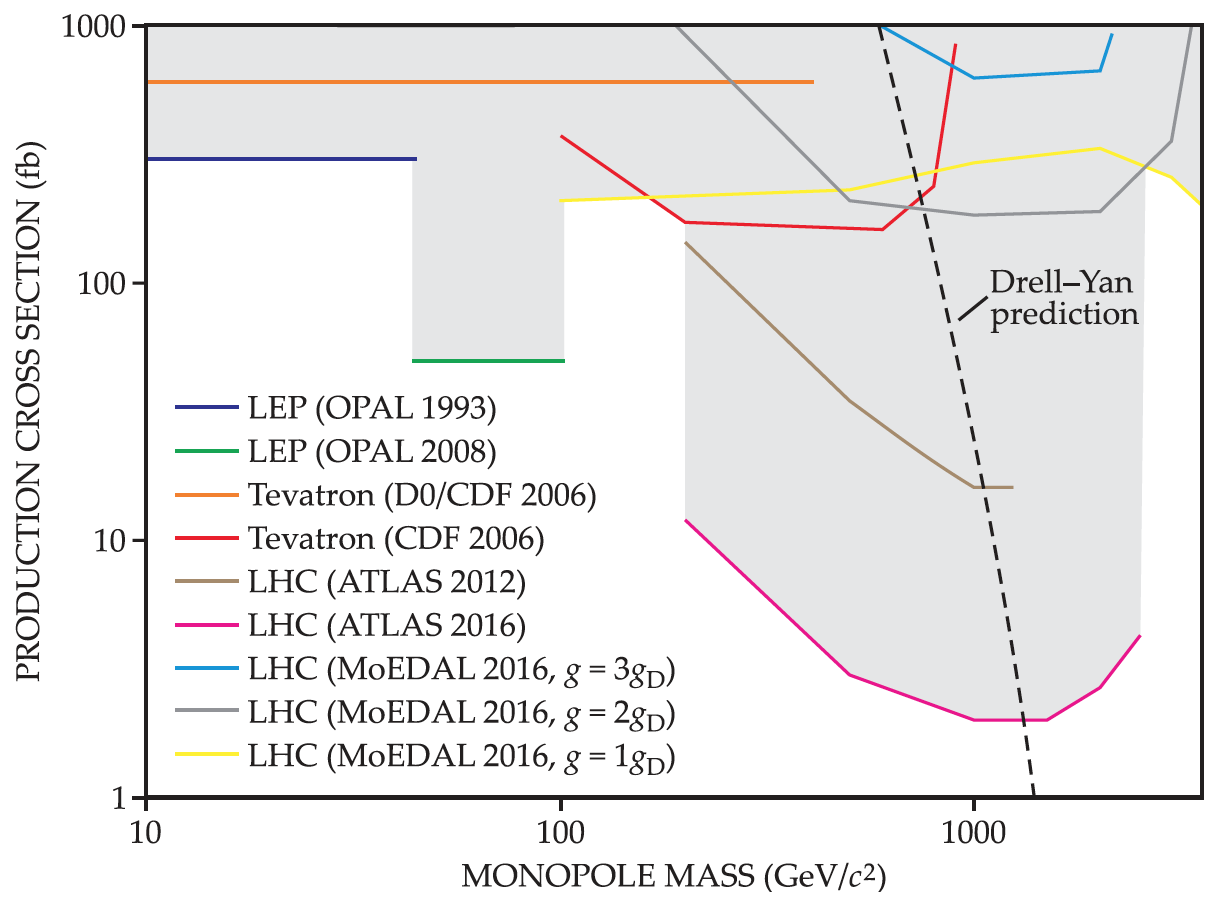}
  \caption{Upper bounds on the monopole production cross sections in various colliders, including MoEDAL. From Ref.~\cite{rajantiept}.}
\label{fig:arttu}
\end{center}
\end{figure}

In the context of MoEDAL searches, 160~kg of prototype MoEDAL trapping detector samples were exposed to 8-TeV proton-proton collisions at the LHC, for an integrated luminosity of 0.75 fb$^{-1}$ during Run~I. No magnetic charge exceeding $0.5g_{\rm D}$ was detected in any of the exposed samples, allowing limits to be placed on monopole production in the mass range 100~GeV$\leq m \leq$ 3500~GeV~\cite{MMT8TeV}. Model-independent cross-section limits have been presented in fiducial regions of monopole energy and direction for $1g_{\rm D}\leq|g|\leq 6g_{\rm D}$, and model-dependent cross-section limits are obtained for Drell-Yan (DY) pair production of spin-1/2 and spin-0 monopoles for $1g_{\rm D}\leq|g|\leq 4g_{\rm D}$ (\emph{cf.} Fig.~\ref{fig:cross_section_limits}). Under the assumption of Drell-Yan cross sections, mass limits are derived for $|g|=2g_{\rm D}$ and $|g|=3g_{\rm D}$ for the first time at the LHC, surpassing the previous result from ATLAS Collaboration~\cite{atlasmono}, which placed limits only for monopoles with magnetic charge $|g|=1 g_{\rm D}$. Caution, however, should be exerted here in the sense that the non-perturbative nature of the large magnetic Dirac charge of the monopole invalidate any perturbative treatment based on Drell-Yan calculations of the pertinent cross sections and hence any result based on the latter is only indicative, due to the lack of any other concrete theoretical treatment. New results with MMTs exposed at 13-TeV $pp$ LHC collisions have been released recently~\cite{MMT13TeV} extending further the limits in terms of monopole mass and charge up to $5g_{\rm D}$ in DY processes. A comparison of the limits on monopole production cross sections set by other colliders with those set by MoeDAL is given in Fig.~\ref{fig:arttu}~\cite{rajantiept}. For more general limits including searches in cosmic radiation, see Ref.~\cite{patrizii}.

A possible explanation for this lack of experimental confirmation of monopoles is Dirac's proposal~\cite{Dirac1931kp,Diracs_idea,khlopov} that monopoles are not seen freely because they form a bound state called \emph{monopolium}~\cite{Monopolium,Monopolium1,Epele0} being confined by strong magnetic forces. Monopolium is a neutral state, hence it is difficult to detect directly at a collider detector, although its decay into two photons would give a rather clear signal for the ATLAS and CMS detectors~\cite{Epele1,Epele2}, which however would not be visible in the MoEDAL detector. Nevertheless according to a novel proposal~\cite{risto}, the LHC radiation detector systems can be used to turn the LHC itself into a new physics search machine by detecting final-state protons $pp\to pXp$ exiting the LHC beam vacuum chamber at locations determined by their fractional momentum losses. Such technique would be appealing for detecting monopolia. Furthermore the monopolium might break up in the medium of MoEDAL into highly-ionising dyons, which subsequently can be detected in MoEDAL~\cite{Acharya:2014nyr}. Moreover its decay via photon emission would produce a peculiar trajectory in the medium, should the decaying states are also magnetic multipoles~\cite{Acharya:2014nyr}.

%%%%%%%%%%%%%%%%%%%%%%%%%%%%%%%%%%%%%%%%%%%%%%%%%%%%%%%%%%%%%%%%%%%%%%%%%%%%%%%%%%%%%%%%%%%%%%%%%%%%
%%%%%%%%%%%%%%%%%%%%%%%%%%%%%%%%%%%%%%%%%%%%%%%%%%%%%%%%%%%%%%%%%%%%%%%%%%%%%%%%%%%%%%%%%%%%%%%%%%%%
\section{Electrically-charged long-lived particles in supersymmetry}\label{sc:susy}

Supersymmetry (SUSY) is an extension of the Standard Model which assigns to each SM field a superpartner field with a spin differing by a half unit. SUSY provides elegant solutions to several open issues in the SM, such as the hierarchy problem, the identity of dark matter, and the grand unification. SUSY scenarios propose a number of massive slowly moving ($\beta \lesssim 0.5$)  electrically charged particles. If they  are sufficiently long-lived to  travel a distance of at least ${\cal O}(1{\rm m})$  before decaying and their $Z/\beta\gtrsim 0.5$,  then they will be detected in the MoEDAL NTDs. No highly-charged particles are expected in such a theory, but there are several scenarios in which supersymmetry may yield massive, long-lived particles that could have electric charges $\pm 1$, potentially detectable in MoEDAL if they are produced with low velocities.

The lightest supersymmetric particle (LSP) is stable in models where $R$~parity is conserved. The LSP should have no strong or electromagnetic interactions, for otherwise it would bind to conventional matter and be detectable in anomalous heavy nuclei~\cite{EHNOS}. Possible weakly-interacting neutral candidates in the Minimal Supersymmetric Standard Model (MSSM) include the sneutrino, which has been excluded by LEP and direct searches, the lightest neutralino $\tilde{\chi}_1^0$ (a mixture of spartners of the $Z, H$ and $\gamma$) and the gravitino $\tilde{G}$.

%%%%%%%%%%%%%%%%%%%%%%%%%%%%%%%%%%%%%%%%%%%%%%%%%%%%%%%%%%%%%%%%%%%%%%%%%%%%%%%%%%%%%%%%%%%%%%%%%%%%
\subsection{Supersymmetric scenarios with $R$-parity violation}

Several scenarios featuring metastable charged sparticles might be detectable in MoEDAL. One such scenario is that $R$~parity {\it may not be exact}, since there is no exact local symmetry associated with either $L$ or $B$, and hence no fundamental reason why they should be conserved. One could consider various ways in which $L$ and/or $B$ could be violated in such a way that $R$ is violated, as represented by the following superpotential terms:
\begin{equation}
W_{RV} \; = \; \lambda^{\prime \prime}_{ijk} {\bar U}_i {\bar D}_j {\bar D}_k
+  \lambda^{\prime}_{ijk} {L}_i {Q}_j {\bar D}_k
+ \lambda_{ijk} {L}_i {L}_j {\bar E}_k
+ \mu_i L_i H,
\label{Rviolation}
\end{equation}
where ${Q}_i, {\bar U}_i, {\bar D}_i, L_i$ and ${\bar E}_i$ denote chiral superfields corresponding to quark doublets, antiquarks, lepton doublets and antileptons, respectively, with $i, j, k$ generation indices. The simultaneous presence of terms of the first and third type in Eq.~(\ref{Rviolation}), namely $\lambda$ and $\lambda^{\prime \prime}$, is severely restricted by lower limits on the proton lifetime, but other combinations are less restricted. The trilinear couplings in Eq.~(\ref{Rviolation}) generate sparticle decays such as ${\tilde q} \to {\bar q} {\bar q}$ or $q \ell$, or ${\tilde \ell} \to \ell \ell$, whereas the bilinear couplings in Eq.~(\ref{Rviolation}) generate Higgs-slepton mixing and thereby also ${\tilde q} \to q \ell$ and ${\tilde \ell} \to \ell \ell$ decays~\cite{Mitsou:2015kpa}. For a nominal sparticle mass $\sim 1$~TeV, the lifetime for such decays would exceed a few nanoseconds for $\lambda,  \lambda^{\prime}, \lambda^{\prime \prime} < 10^{-8}$. 

If $R$~parity is broken, the LSP would be unstable, and might be charged and/or coloured. In the former case, it might be detectable directly at the LHC as a massive slowly-moving charged particle. In the latter case, the LSP would bind with light quarks and/or gluons to make colour-singlet states, the so-called \emph{R-hadrons}, and any charged state could again be detectable as a massive slowly-moving charged particle. If $\lambda \ne 0$, the prospective experimental signature would be similar to a stau next-to-lightest sparticle (NLSP) case to be discussed later. On the other hand, if $\lambda^{\prime}$ or $\lambda^{\prime \prime} \ne 0$, the prospective experimental signature would be similar to a stop NLSP case, yielding the possibility of charge-changing interactions while passing through matter. This could yield  a metastable charged particle, created whilst passing through the material surrounding the intersection point,  that would be detected by MoEDAL. 

%%%%%%%%%%%%%%%%%%%%%%%%%%%%%%%%%%%%%%%%%%%%%%%%%%%%%%%%%%%%%%%%%%%%%%%%%%%%%%%%%%%%%%%%%%%%%%%%%%%%
\subsection{Metastable lepton NLSP in the CMSSM with a neutralino LSP}

However, even if $R$~parity {\it is} exact, the NLSP may be long lived. This would occur, for example, if the LSP is the gravitino, or if the mass difference between the NLSP and the neutralino LSP is small, offering more scenarios for long-lived charged sparticles. In {\it neutralino dark matter} scenarios based on the constrained MSSM (CMSSM), for instance, the most natural candidate for the NLSP is the lighter stau slepton ${\tilde \tau_1}$~\cite{stauNLSP}, which could be long lived if $m_{\tilde \tau_1} - m_{\tilde{\chi}_1^0}$ is small. There are several regions of the CMSSM parameter space that are compatible with the constraints imposed by unsuccessful searches for sparticles at the LHC, as well as the discovered Higgs boson mass. These include a strip in the focus-point region where the relic density of the LSP is brought down into the range allowed by cosmology because of its relatively large Higgsino component, a region where the relic density is controlled by rapid annihilation through direct-channel heavy Higgs resonances, and a strip where the relic LSP density is reduced by coannihilations with near-degenerate staus and other sleptons. It was found in a global analysis that the two latter possibilities are favoured~\cite{MC8}.

\begin{figure}[ht]
\centering\sidecaption
\includegraphics[width=0.43\textwidth]{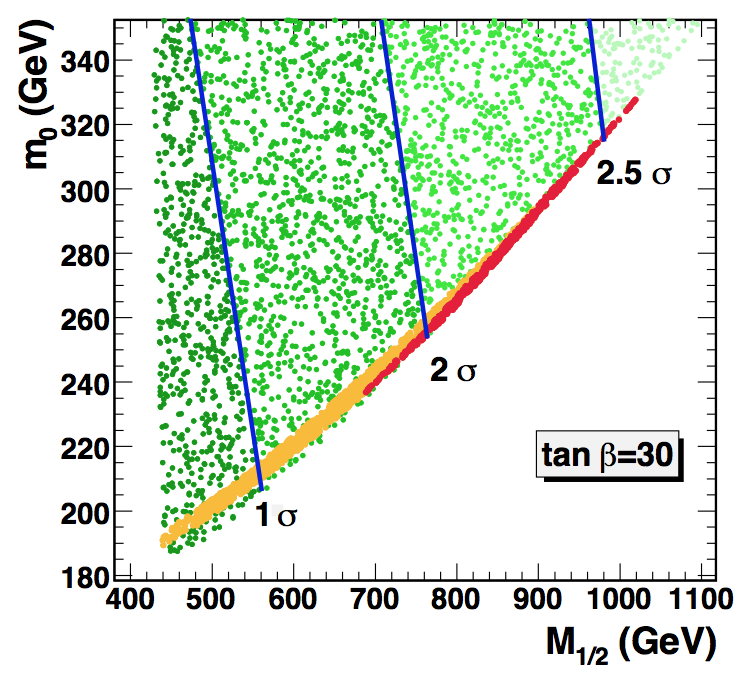}
\caption{Allowed parameter regions in $M_{1/2} - m_0$ plane fixing $A_0=600$~GeV and $\tan\beta=30$. The red (dark) narrow band is consistent region with dark matter abundance and $\delta m < m_\tau$ and the yellow (light) narrow band is that with $\delta m > m_\tau$. The green regions are inconsistent with the dark matter abundance, and in the white (excluded) area the LSP is the stau. The favoured regions of the muon anomalous magnetic moment at $1\sigma$, $2\sigma$ and $2.5\sigma$ confidence level are indicated by solid lines. From Ref.~\cite{oscar}.}
\label{fg:oscar}
\end{figure}

In the coannihilation region of the CMSSM, the lighter ${\tilde \tau_1}$ is expected to be the lightest slepton~\cite{stauNLSP}, and the $\tilde\tau_1-\tilde{\chi}_1^0$ mass difference may well be smaller than $m_\tau$: indeed, this is required at large LSP masses. In this case, the dominant stau decays for $m_{\tilde \tau_1} - m_{\tilde{\chi}_1^0} > 160$~MeV are expected to be into three particles: $\tilde{\chi}_1^0 \nu \pi$ or $\tilde{\chi}_1^0 \nu \rho$. If $m_{\tilde \tau_1} - m_{\tilde{\chi}_1^0} < 1.2$~GeV, the ${\tilde \tau_1}$ lifetime is calculated to be so long, in excess of $\sim 100$~ns, that it is likely to escape the detector before decaying, and hence would be detectable as a massive, slowly-moving charged particle~\cite{Sato,oscar}. The relevance of such scenarios while considering cosmological constraints is demonstrated in Fig.~\ref{fg:oscar}. Even is lepton-flavor violating couplings are allowed, the long lifetime of the staus remains~\cite{oscar}.

%%%%%%%%%%%%%%%%%%%%%%%%%%%%%%%%%%%%%%%%%%%%%%%%%%%%%%%%%%%%%%%%%%%%%%%%%%%%%%%%%%%%%%%%%%%%%%%%%%%%
\subsection{Metastable sleptons in gravitino LSP scenarios}

On the other hand, in {\it gravitino dark matter} scenarios with more general options for the pattern of supersymmetry breaking, other options appear quite naturally, including the lighter selectron or smuon, or a sneutrino~\cite{sleptonNLSP}, or the lighter stop squark ${\tilde t_1}$~\cite{stopNLSP}. If the gravitino ${\tilde G}$ is the LSP, the decay rate of a slepton NLSP is given by 
\begin{equation}
\Gamma ( {\tilde \ell} \to {\tilde G} \ell) = \dfrac{1}{48 \pi M_*^2} \dfrac{m_{\tilde \ell}^5}{M_{\tilde G}^2}
\left[ 1 - \dfrac{M_{\tilde G}^2}{m_{\tilde \ell}^2} \right]^{4},
\label{telldecay}
\end{equation}
where $M_*$ is the Planck scale. Since $M_*$ is much larger than the electroweak scale, the NLSP lifetime is naturally very long.   

Gravitino (or axino) LSP with a long-lived charged stau may arise in gauge mediation and minimal supergravity models~\cite{Nojiri}. Large part of the parameter space potentially attractive for long-lived slepton searches with MoEDAL are compatible with cosmological constraints on the dark-matter abundance in superweakly interacting massive particle scenarios~\cite{Feng}.

%%%%%%%%%%%%%%%%%%%%%%%%%%%%%%%%%%%%%%%%%%%%%%%%%%%%%%%%%%%%%%%%%%%%%%%%%%%%%%%%%%%%%%%%%%%%%%%%%%%%
\subsection{Long-lived gluinos in split supersymmetry}

The above discussion has been in the context of the CMSSM and similar scenarios where all the supersymmetric partners of Standard Model particles have masses in the TeV range. Another scenario is ``split supersymmetry'', in which the supersymmetric partners of quarks and leptons are very heavy, of a scale $m_s$, whilst the supersymmetric partners of SM bosons are relatively light~\cite{splitSUSY}. In such a case, the gluino could have a mass in the TeV range and hence be accessible to the LHC, but would have a very long lifetime:
\begin{equation}
\tau \approx 8 \left( \dfrac{m_s}{10^9~{\rm GeV}} \right)^4 \left( \dfrac{1~{\rm TeV}}{m_{\tilde{g}}} \right)^5~{\rm s}.
\label{gluinotau}
\end{equation}
Long-lived gluinos would form long-lived gluino R-hadrons including gluino-gluon \emph{(gluinoball)} combinations, gluino-$q{\bar q}$ \emph{(mesino)} combinations and gluino-$qqq$ \emph{(baryino)} combinations. The heavier gluino hadrons would be expected to decay into the lightest species, which would be metastable, with a lifetime given by Eq.~(\ref{gluinotau}), and it is possible that this metastable gluino hadron could be charged.

In the same way as stop hadrons, gluino hadrons may flip charge through conventional strong interactions as they pass through matter, and it is possible that one may pass through most of a conventional LHC tracking detector undetected in a neutral state before converting into a metastable charged state that could be detected by MoEDAL. 

%%%%%%%%%%%%%%%%%%%%%%%%%%%%%%%%%%%%%%%%%%%%%%%%%%%%%%%%%%%%%%%%%%%%%%%%%%%%%%%%%%%%%%%%%%%%%%%%%%%%
\subsection{Experimental considerations} 

There are several considerations supporting the complementary aspects of MoEDAL w.r.t.\ ATLAS and CMS when discussing the observability of (meta-)stable massive electrically-charged particles. Most of them stem from MoEDAL being ``time-agnostic'' due to the passive nature of its detectors. Therefore signal from very slowly moving particles will not be lost due to arriving in several consecutive bunch crossings. ATLAS and CMS, on the other hand, perform triggered-based analyses relying either on triggering on accompanying ``objects'', e.g.\ missing transverse energy, or by developing and deploying specialised triggers. In both cases the efficiency may lower and in the former the probed parameter space may be reduced. MoEDAL is mainly limited by the geometrical acceptance of the detectors, especially the MMTs, and by the requirement of passing the $Z/\beta$ threshold of NTDs. In general ATLAS and CMS have demonstrated to cover high-velocities $\beta \gtrsim 0.2$, while MoEDAL is sensitive to lower ones $\beta \lesssim 0.2$. 

When discussing the detection of particles stopped \emph{(trapped)} in material that they may decay later, different possibilities are explored. CMS and ATLAS look in empty bunch crossings for decays of trapped particles into jets. MoEDAL MMTs may be monitored in a underground/basement laboratory for tracks arising from such decays. The background in the latter case, coming from cosmic rays, should be easier to control and assess. The probed lifetimes should be larger due to the unlimited monitoring time.

%%%%%%%%%%%%%%%%%%%%%%%%%%%%%%%%%%%%%%%%%%%%%%%%%%%%%%%%%%%%%%%%%%%%%%%%%%%%%%%%%%%%%%%%%%%%%%%%%%%%
%%%%%%%%%%%%%%%%%%%%%%%%%%%%%%%%%%%%%%%%%%%%%%%%%%%%%%%%%%%%%%%%%%%%%%%%%%%%%%%%%%%%%%%%%%%%%%%%%%%%
\section{Doubly-charged Higgs bosons}\label{sc:lrsm} 

Doubly-charged particles appear in many theoretical scenarios beyond the SM. For example, doubly-charged scalar states, usually termed doubly-charged Higgs fields, appear in left-right symmetric models~ \cite{Pati1974yy,LRSM,LRSMa} and in see-saw models for neutrino masses with Higgs triplets. A number of models encompasses additional symmetries and extend the SM Higgs sector by introducing doubly-charged Higgs bosons. A representative example of such a model is the L-R Symmetric Model (LRSM)~\cite{Pati1974yy,LRSM,LRSMa}, proposed to amend the fact that the weak-interaction couplings are strictly left handed by extending the gauge group of the SM so as to include a right-handed sector. The simplest realisation is an LRSM~\cite{Pati1974yy, LRSM}  that postulates  a right-handed version of the weak interaction, whose gauge symmetry is spontaneously broken at high mass scale, leading to the parity-violating  SM. This model naturally accommodates recent data on neutrino oscillations and the existence of small neutrino masses. The model generally requires Higgs triplets containing doubly-charged Higgs bosons ($H^{\pm\pm}$)  $\Delta_{R}^{++}$ and $\Delta_{L}^{++}$, which could be light in the minimal supersymmetric left-right model~\cite{LRSUSY}.

Single production of a doubly-charged Higgs boson at the LHC proceeds via vector boson fusion, or through the fusion of a singly-charged Higgs boson with either a $W^\pm$ or another singly-charged Higgs boson. The amplitudes of the  $W_{L} W_{L}$ and $W_{R} W_{R}$ vector boson fusion processes are proportional to $v_{L,R}$, the vacuum expectation values of the neutral members of the scalar triplets of the  LRSM. For the case of $\Delta_{R}^{++}$ production, the vector boson fusion process dominates. Pair production of doubly-charged Higgs bosons is also possible via a Drell-Yan process, with $\gamma$, $Z$ or $Z_{R}$ exchanged in the $s$-channel, but at a high kinematic price since substantial energy is required to produce two heavy particles. In the case of $\Delta_{L}^{++}$, double production may nevertheless be the only possibility if $v_{L}$ is very small or vanishing.

\begin{figure}[ht]
\centering
\sidecaption
\includegraphics[width=0.42\textwidth]{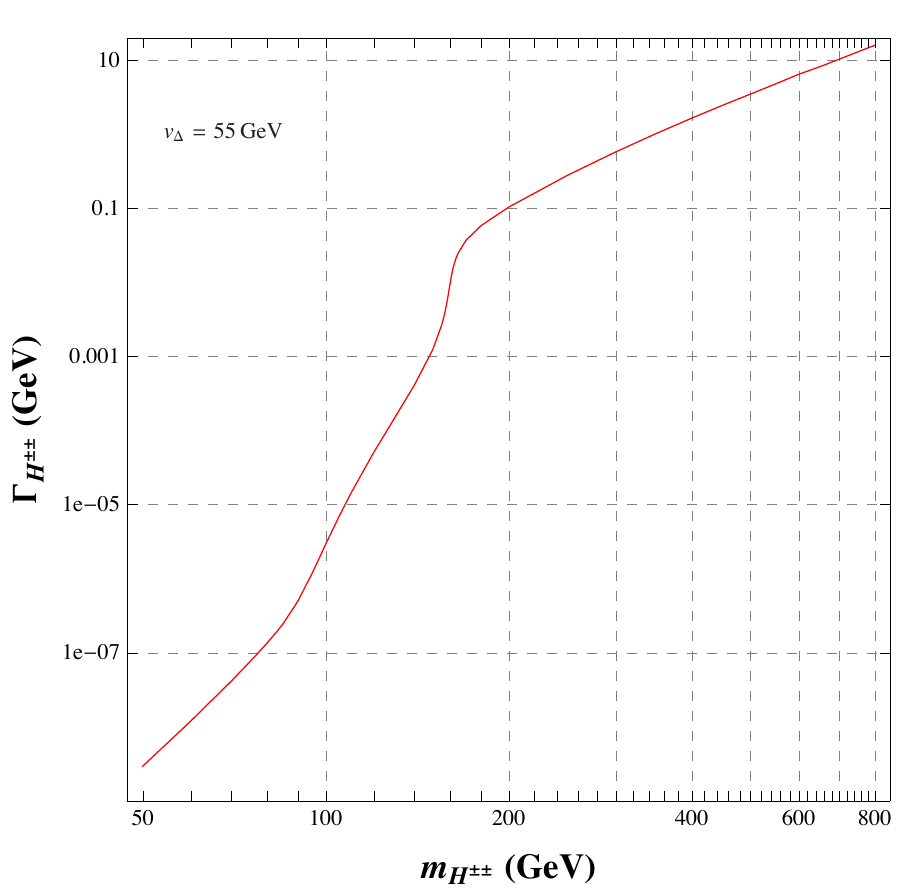}
\caption{Partial decay width of $H^{\pm \pm} \rightarrow W^{\pm} W^{\pm}$ as a function of $m_{H^{\pm\pm}}$ for $v_{\Delta} = 55~{\rm GeV}$. From Ref.~\cite{Chiang:2012dk}.} \label{fig:width}
\end{figure}

The decay of a doubly-charged Higgs boson can proceed via several channels. The dilepton signature leads to the (experimentally clean) final state $q\bar{q} \rightarrow \Delta^{++}_L\Delta^{--}_L \rightarrow 4\ell$. However as long as the triplet vacuum expectation value, $v_\Delta$, is much larger than $10^{-4}~{\rm GeV}$, the doubly-charged Higgs decay predominantly into a pair of same-sign $W$ bosons. For very small Yukawa couplings $H_{\ell\ell} \lesssim  10^{-8}$, the doubly-charged Higgs boson can be quasi-stable~\cite{Chiang:2012dk}. In Fig.~\ref{fig:width}, the partial decay width of the doubly charged Higgs boson into a $W$ boson pair is shown as a function of its mass. For $v_\Delta \gg 10^{-4}~{\rm GeV}$, this partial width is roughly equal to the total width of the doubly charged Higgs boson. In the case of long lifetimes, slowly moving pseudo-stable Higgs bosons could be detected in the MoEDAL NTDs. For example with CR39, one could detect doubly-charged Higgs particles moving with speeds less than around $\beta \simeq 0.4$. 

%%%%%%%%%%%%%%%%%%%%%%%%%%%%%%%%%%%%%%%%%%%%%%%%%%%%%%%%%%%%%%%%%%%%%%%%%%%%%%%%%%%%%%%%%%%%%%%%%%%%
\section{Black hole remnants in large extra dimensions}\label{sc:bh}

Over the last decades, models based on compactified extra spatial dimensions (ED) have been proposed in order to explain the large gap between the electroweak (EW) and the Planck scale of $M_{\rm EW}/M_{\rm PL}  \approx 10^{-17}$. The four main scenarios relevant for searches at LHC the Arkani-Hamed-Dimopoulos-Dvali (ADD) model of large extra dimensions~\cite{ADD}, the Randall-Sundrum (RS) model of warped extra dimensions~\cite{Randall}, TeV$^{-1}$-sized extra dimensions~\cite{TEV-1}, and the Universal Extra Dimensions (UED) model~\cite{UED}.

The existence of extra spatial dimensions~\cite{ADD,Randall} and a sufficiently small fundamental scale of gravity open up the possibility that microscopic black holes be produced and detected~\cite{ED1, bhevaporation, fischler1, CHARYBDIS, ED2,ED3}  at the LHC. Once produced, the black holes will undergo an evaporation process categorised in three stages~\cite{bhevaporation, fischler1}: the \emph{balding phase}, the actual \emph{evaporation phase}, and finally   the Planck phase. It is generally assumed that the black hole will decay completely to some last few SM particles. However, another intriguing possibility is that the remaining energy is carried away by a stable remnant.

The prospect of microscopic black hole production at the LHC within the framework of models with large extra dimensions has been studied in Ref.~\cite{ADD}. Black holes produced at the LHC are expected to decay with an average multiplicity of $\sim10-25$ into SM particles,  most of which will be charged, though the details of the multiplicity distribution depend on the number of extra dimensions~\cite{BHMULT}. After the black holes have evaporated off enough energy to reach the remnant mass, some will have accumulated a net electric charge. According to purely statistical considerations, the probability for being left with highly-charged black hole remnants drops fast with the deviation from the average. The largest fraction of the black holes should have charges $\pm1$ or zero, although a smaller but non-negligible fraction would be multiply charged.

The fraction of charged black-hole remnants has been estimated~\cite{BHMULT,Hossenfelder:2005ku} using the {\tt PYTHIA} event generator~\cite{PYTHIA} and the {\tt CHARYBDIS} program~\cite{CHARYBDIS}. It was  assumed that the effective temperature of the black hole drops towards zero for a finite remnant mass, $M_{R}$. The value of $M_{R}$ does not noticeably affect the investigated charge distribution, as it results from the very general statistical distribution of the charge of the emitted particles.

\begin{figure}[htb]
\centering
\sidecaption
\includegraphics[width=0.42\textwidth]{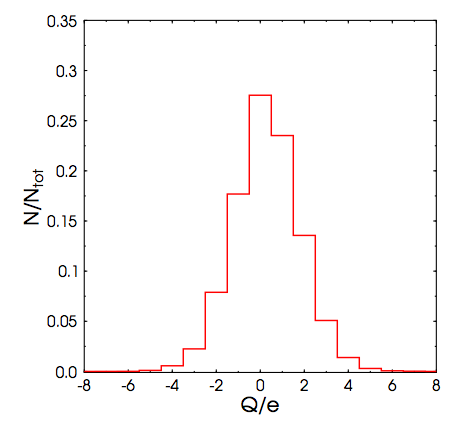}
\caption{The distribution of black-hole remnant charges in proton-proton interactions at $\sqrt{s} = 14~{\rm TeV}$ calculated with the
{\tt PYTHIA} event generator~\cite{PYTHIA} and the {\tt CHARYBDIS} program~\cite{CHARYBDIS}. From Ref.~\cite{Hossenfelder:2005ku}.} \label{Fig:Qofremnants}
\end{figure}

Thus, independent of the underlying quantum-gravitational  assumption leading to the remnant formation, it was found that about 30\% of the remnants are neutral, whereas $\sim$ 40\% would be singly-charged black holes, and the  remaining $\sim$30\% of remnants would be multiply-charged.  The distribution of the  remnant charges obtained  is shown in Fig.~\ref{Fig:Qofremnants}. The black hole remnants  considered here are heavy, with masses of a TeV or more. A significant  fraction of the black-hole remnants produced would have a  Z/$\beta$ of greater than five, high enough to register in the CR39 NTDs forming the LT-NTD  sub-detector of MoEDAL.

%%%%%%%%%%%%%%%%%%%%%%%%%%%%%%%%%%%%%%%%%%%%%%%%%%%%%%%%%%%%%%%%%%%%%%%%%%%%%%%%%%%%%%%%%%%%%%%%%%%%
%%%%%%%%%%%%%%%%%%%%%%%%%%%%%%%%%%%%%%%%%%%%%%%%%%%%%%%%%%%%%%%%%%%%%%%%%%%%%%%%%%%%%%%%%%%%%%%%%%%%
\section{D-matter}\label{sc:dmatter}

Some versions of string theory include higher-dimensional ``domain-wall''-like membrane \emph{(brane)} structures in space-time, called \emph{D-branes}. In some cases the bulk is ``punctured'' by lower-dimensional D-brane defects, which are either point-like or have their longitudinal dimensions compactified~\cite{westmuckett}. From a low-energy observer's perspective, such structures would effectively appear to be point-like \emph{D-particles}. The latter have  dynamical degrees of freedom, thus they can be treated as quantum  excitations above the vacuum~\cite{westmuckett,shiu} collectively referred to as {\it D-matter}. D-matter states are non-perturbative stringy objects with masses of order $m_D \sim M_s/g_s$, where $g_s $ is the string coupling, typically of order one so that the observed gauge and gravitational couplings is reproduced. Hence, the D-matter states could be light enough to be phenomenologically relevant at the LHC.

Depending on their type, D-branes could carry integral or torsion (discrete) charges with the lightest D-particle (LDP) being stable. Therefore the LDPs are possible candidates for cold dark matter~\cite{shiu}. D-particles are solitonic non-perturbative objects in the string/brane theory. As discussed in the relevant literature~\cite{shiu}, there are similarities and differences between D-particles and magnetic monopoles with non-trivial cosmological implications~\cite{Witten2002wb,westmuckett,Mavromatos:2010jt,mitsou}. An important difference is that they could have {\it perturbative} couplings, with no magnetic charge in general. Nonetheless, in the context of brane-inspired gauge theories, brane states with magnetic charges can be constructed, which would manifest themselves in MoEDAL in a manner similar to  magnetic monopoles. 
 
\begin{figure}[ht]
\centering
\includegraphics[width=0.3\textwidth]{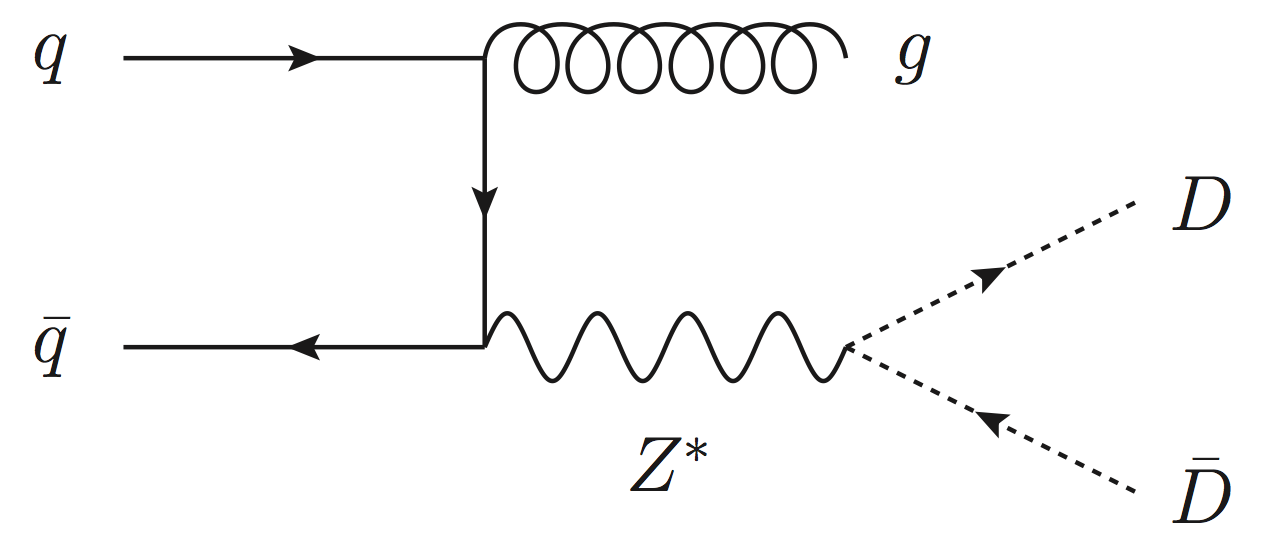}
\caption{An example of parton-level diagrams for production of D-particles by $q\bar{q}$ collisions in a generic D-matter low-energy model~\cite{mitsou}. }
\label{fig:dproduction}
\end{figure}

Non-magnetically-charged D-matter, on the other hand, could be produced at colliders and also produce interesting signals of direct relevance to the MoEDAL experiment. For instance, excited states of D-matter (${\rm D}^\star$) can be electrically-charged. For typical string couplings of phenomenological relevance, the first few massive levels may be accessible to the LHC.  Depending on the details of the microscopic model considered, and the way the SM is embedded, such massive charged states can be relatively long-lived, and could likewise be detectable with MoEDAL. D-matter/antimatter pairs can be produced~\cite{Mavromatos:2010jt,mitsou} by the decay of intermediate off-shell $Z$-bosons, as shown in Fig.~\ref{fig:dproduction}. 

%%%%%%%%%%%%%%%%%%%%%%%%%%%%%%%%%%%%%%%%%%%%%%%%%%%%%%%%%%%%%%%%%%%%%%%%%%%%%%%%%%%%%%%%%%%%%%%%%%%%
\section{Summary and outlook}\label{sc:summary}

MoEDAL is going to extend considerably the LHC reach in the search for (meta)stable highly ionising particles. The latter are predicted in a variety of theoretical models and include: magnetic monopoles, SUSY stable (or, rather, long-lived) spartners, quirks, strangelets, Q-balls, fractionally-charged massive particles, etc~\cite{Acharya:2014nyr}. Such particles can be light enough to be producible at the LHC energies (see e.g. Q-balls in the context of some SUSY or brane models~\cite{kehagias}).
In the talk we have described searches for only a subset of those particles, due to lack of space. Specifically, we discussed monopoles, partners in some SUSY models, doubly charged Higgs bosons, black hole remnants in models with large extra spatial dimensions, as well as some (more exotic) scenarios on D-matter which characterises some brane models. 

The MoEDAL design is optimised to probe precisely all such long lived states, unlike the other LHC experiments~\cite{DeRoeck:2011aa}. Furthermore it combines different detector technologies: plastic nuclear track detectors (NTDs), trapping volumes and pixel sensors~\cite{moedal-tdr}. The first physics results, pertaining to magnetic monopole trapping detectors, obtained with LHC Run~I data, have been published~\cite{MMT8TeV} and the corresponding analysis at 13~TeV has been released recently~\cite{MMT13TeV}. The MoEDAL Collaboration is preparing new analyses with more Run~II data, with other detectors (NTDs) and with a large variety of interpretations involving not only magnetic but also electric charges.

%%%%%%%%%%%%%%%%%%%%%%%%%%%%%%%%%%%%%%%%%%%%%%%%%%%%%%%%%%%%%%%%%%%%%%%%%%%%%%%%%%%%%%%%%%%%%%%%%%%%
%%%%%%%%%%%%%%%%%%%%%%%%%%%%%%%%%%%%%%%%%%%%%%%%%%%%%%%%%%%%%%%%%%%%%%%%%%%%%%%%%%%%%%%%%%%%%%%%%%%%
\section*{Acknowledgements}

The authors are grateful to the ICNFP2016 organisers for scheduling the ``Mini-Workshop on MoEDAL'' during the Conference and for the kind invitation to present these talks in it. The work of NEM is supported in part by the UK Science and Technology Facilities Council (STFC) via the grants ST/L000326/1 and ST/P000258/1. VAM acknowledges support by the Spanish Ministry of Economy and Competitiveness (MINECO) under the project FPA2015-65652-C4-1-R, by the Generalitat Valenciana through the project PROMETEO~II/2013-017 and by the Spanish National Research Council (CSIC) under the CT Incorporation Program 201650I002. 

%%%%%%%%%%%%%%%%%%%%%%%%%%%%%%%%%%%%%%%%%%%%%%%%%%%%%%%%%%%%%%%%%%%%%%%%%%%%%%%%%%%%%%%%%%%%%%%%%%%%
%%%%%%%%%%%%%%%%%%%%%%%%%%%%%%%%%%%%%%%%%%%%%%%%%%%%%%%%%%%%%%%%%%%%%%%%%%%%%%%%%%%%%%%%%%%%%%%%%%%%

\end{document}